\documentclass[twocolumn,twoside,slac_two]{revtex4}
\usepackage{graphicx}
\usepackage{fancyhdr}
\begin{document}

\title{MAGIC Telescopes observation of the BL Lac objects 1ES~1215+303 and 1ES~1218+304}
\author{S.~Lombardi}
\affiliation{Universit\`{a} di Padova and INFN, I-35131 Padova, Italy}
\author{E.~Lindfors}
\affiliation{Tuorla Observatory, Department of Physics and Astronomy, University of Turku, FI-21500 Piikki\"{o}, Finland}
\author{J.~Becerra Gonz\'{a}lez}
\affiliation{Instituto de Astrof\'{i}sica de Canarias, E-38200 La Laguna, Tenerife, Spain}
\author{P.~Colin}
\affiliation{Max-Planck-Institut f\"ur Physik, D-80805 M\"unchen, Germany}
\author{J.~Sitarek}
\affiliation{IFAE, Edifici Cn., Campus UAB, E-08193 Bellaterra, Spain}
\author{A.~Stamerra}
\affiliation{Universit\`a  di Siena, and INFN Pisa, I-53100 Siena, Italy}
\author{on behalf of the MAGIC Collaboration}
\affiliation{}
\begin{abstract}
The two BL Lac objects, 1ES~1215+303 and 1ES~1218+304, separated by
$0.8^{\circ}$, were observed with the MAGIC Cherenkov telescopes in
2010--2011. The January 2011 observations resulted in the first
detection above $100$~GeV of 1ES~1215+303 (known also as ON-325) which
has been flagged as a promising Very High Energy (VHE, $E>100$~GeV)
$\gamma$-ray source candidate by the \emph{Fermi}--LAT collaboration
in October 2010. The January 2011 observations were triggered by the
high optical state of the source, as reported by the Tuorla blazar
monitoring program. A comparison with the 2010 data suggests that
1ES~1215+303 was also flaring in VHE $\gamma$-rays. In addition, the
Swift Target of Opportunity (ToO) observations in X-rays showed that
the flux was almost doubled with respect to previous observations
(December 2009). Instead, 1ES~1218+304 is a well known VHE
$\gamma$-ray emitter located in the same field of view, which was then
simultaneously observed with MAGIC. In this contribution we present
preliminary results of the MAGIC observations of these two VHE
$\gamma$-ray emitting AGNs.
\end{abstract}

\maketitle

\thispagestyle{fancy}


\section{Introduction}
\subsection{The MAGIC Telescopes}
\begin{figure}
\includegraphics[width=80mm]{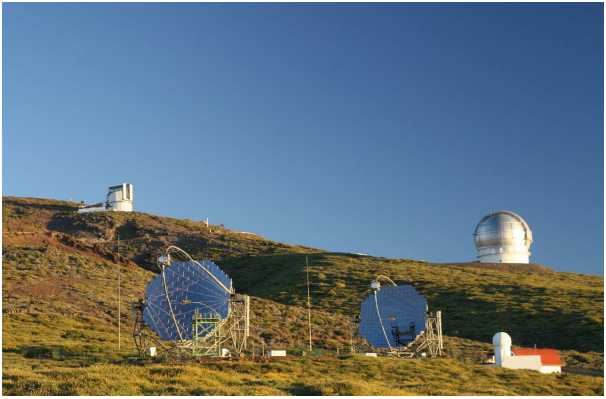}
\caption{The MAGIC Telescopes: MAGIC-I on the left and MAGIC-II on 
the right. Image credit Robert Wagner.}
\label{MAGIC}
\end{figure}

MAGIC consists of two 17 m dish Imaging Air Cherenkov Telescopes located 
in the Canary Island of La Palma, at 2200~m a.s.l. (Figure~\ref{MAGIC}). 
The stereoscopic system has been in operation since fall 2009 and has 
a sensitivity of $0.8\%$ Crab Nebula flux above $\sim 300$~GeV in $50$~h of 
observations, and a trigger threshold of $50$~GeV, which is the 
lowest among the existing IACTs. The MAGIC cameras have a field of view 
of $3.5^{\circ}$. Details on the performance of the MAGIC stereoscopic 
system are presented in~\cite{performance}.
\subsection{Optically Triggered Target of Opportunity Observations}
\begin{figure}
\includegraphics[width=50mm,angle=270]{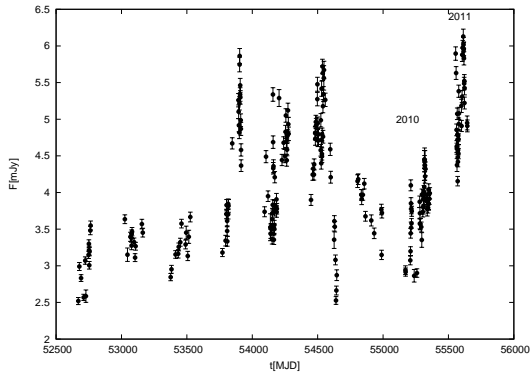}
\caption{Long-term optical R-band light curve of 1ES~1215+303 from the Tuorla 
blazar monitoring program. The outburst in January 2011 triggered the target 
of opportunity observations with MAGIC.}
\label{opticalLC1215}
\end{figure}

MAGIC has been successfully performing optically triggered target of
opportunity (ToO) observations of active galactic nuclei (AGN) since 
the beginning of its science observations. The optical triggers have 
been provided by the Tuorla blazar monitoring program\footnote{http://users.utu.fi/kani/1m}. 
The observations are performed with the KVA~35~cm telescope located on 
La Palma, but remotely operated from Tuorla Observatory. The long-term
monitoring program consists of $>50$ blazars that are considered to be
good candidates to emit Very High Energy (VHE, $E>100$~GeV) $\gamma$-rays 
based on their X-ray and $\gamma$-ray properties.\\
The ToO observations with MAGIC have resulted so far in the discovery 
of many new VHE $\gamma$-ray emitting sources, most recently B3~2247+381 
(ATel$\#$2910) and 1ES~1215+303 (ATel$\#$3100, this contribution). However, 
in many cases it has not been possible to confirm if the sources were 
in a high VHE $\gamma$-ray state during the observations and therefore 
the connection between the optical and VHE $\gamma$-ray states has 
remained an open question. \\
In the first days of January 2011 1ES~1215+303 was observed to be in a 
high optical state (see Figure~\ref{opticalLC1215}) which triggered the 
MAGIC observations of the source. Here we present preliminary results 
of these observations together with previous observations carried out 
in 2010, where the source was in a lower optical state.
\subsection{BL Lac Objects 1ES~1215+303 and 1ES~1218+304}
BL Lac objects belong to a subclass of AGN where the relativistic
jet is pointing very close to our line of sight, causing flux
enhancement and fast variability in all wavebands. Their spectral
energy distribution (SED) is characterized by a typical double bump
shape. The first and second bumps are associated respectively to
synchrotron and inverse Compton emissions.\\
1ES~1215+303 (also known as ON~325) is a high energy peaking BL~Lac object. Two values can be found 
for its redshift in the literature: z=0.130, and
z=0.237\footnote{http://ned.ipac.caltech.edu/}. However, since none of
the references given for the redshift show the optical spectra, it is
difficult to judge which is the most correct value. Nevertheless,
1ES~1215+303 has a bright host galaxy of magnitude $\mbox{R}=16.24$ (see
e.g.~\cite{2003A&A...400...95N}), which can be used for estimating the
redshift. Following the methodology of \cite{2008A&A...487L..29N} we
derive z=0.13$\pm0.04$ (Nilsson, 2011, priv. comm.),  favoring the
lower redshift value.\\ 
1ES~1215+303 was classified as promising
candidate TeV blazar in~\cite{CG} and has been observed several times
in VHE $\gamma$-rays prior to the observations presented here,
providing only upper limits (Whipple:
F$(>430$~GeV$)<1.89\cdot10^{-11}$cm$^{-2}$ s$^{-1}$~\cite{Horan},
MAGIC: F$(>120$~GeV$)<3.5\cdot10^{-11}$ cm$^{-2}$
s$^{-1}$~\cite{stacked}).  The source was also present in the
\emph{Fermi}--LAT bright AGN catalog~\cite{brightlist}, showing
variable flux and a hard spectrum ($\Gamma=-1.89\pm0.06$). In the
\emph{Fermi}--LAT band, 1ES~1215+303 is an exceptional source. The
source underwent a large outburst in late 2008 and in the first bright AGN
catalog~\cite{fermi_lightcurves} it was the only high energy peaking
source that showed significant variability. 1ES~1215+303 has also been
flagged as a promising VHE $\gamma$-ray source candidate by the
\emph{Fermi}--LAT collaboration in October 2010.\\
1ES~1218+304 is another high-peaking BL Lac object located only $0.8^{\circ}$ away
from 1ES~1215+303. It has a redshift of 0.182 and was first detected
to emit VHE $\gamma$-rays by MAGIC in 2005~\cite{1218}. In 2009
VERITAS reported fast variability from the source, the peak flux
reaching $\sim$20$\%$ of the Crab Nebula flux~\cite{1218_ver}.  In the
\emph{Fermi}--LAT one year catalog~\cite{1FGL} the source is flagged
non-variable. Since the measured VHE spectrum of 1ES~1218+304 is
particularly hard for its redshift, an intrinsic SED with an inverse
Compton peak above 1 TeV is expected. The $\gamma$-ray emission is
thus strongly interacting with the extragalactic backgroung light
(EBL) before reaching us, making the source a good candidate to probe
the EBL~\cite{al06,ac09} or the extragalactic magnetic field
\cite{ne10}.
\section{Observations and Preliminary Results}
\begin{figure}
\includegraphics[width=80mm]{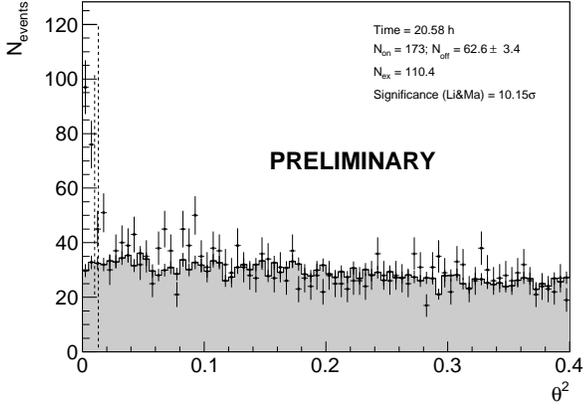}
\caption{Distribution of the square distances ($\theta^2$, in degree$^2$) between the reconstructed 
shower directions and the 1ES~1215+303 position, for data taken in January-February 2011 
with MAGIC. The grey filled histogram represents the expected background estimated with 5~Off 
positions at the same distance from the camera center.}
\label{theta2}
\end{figure}
\begin{figure}
\includegraphics[width=80mm]{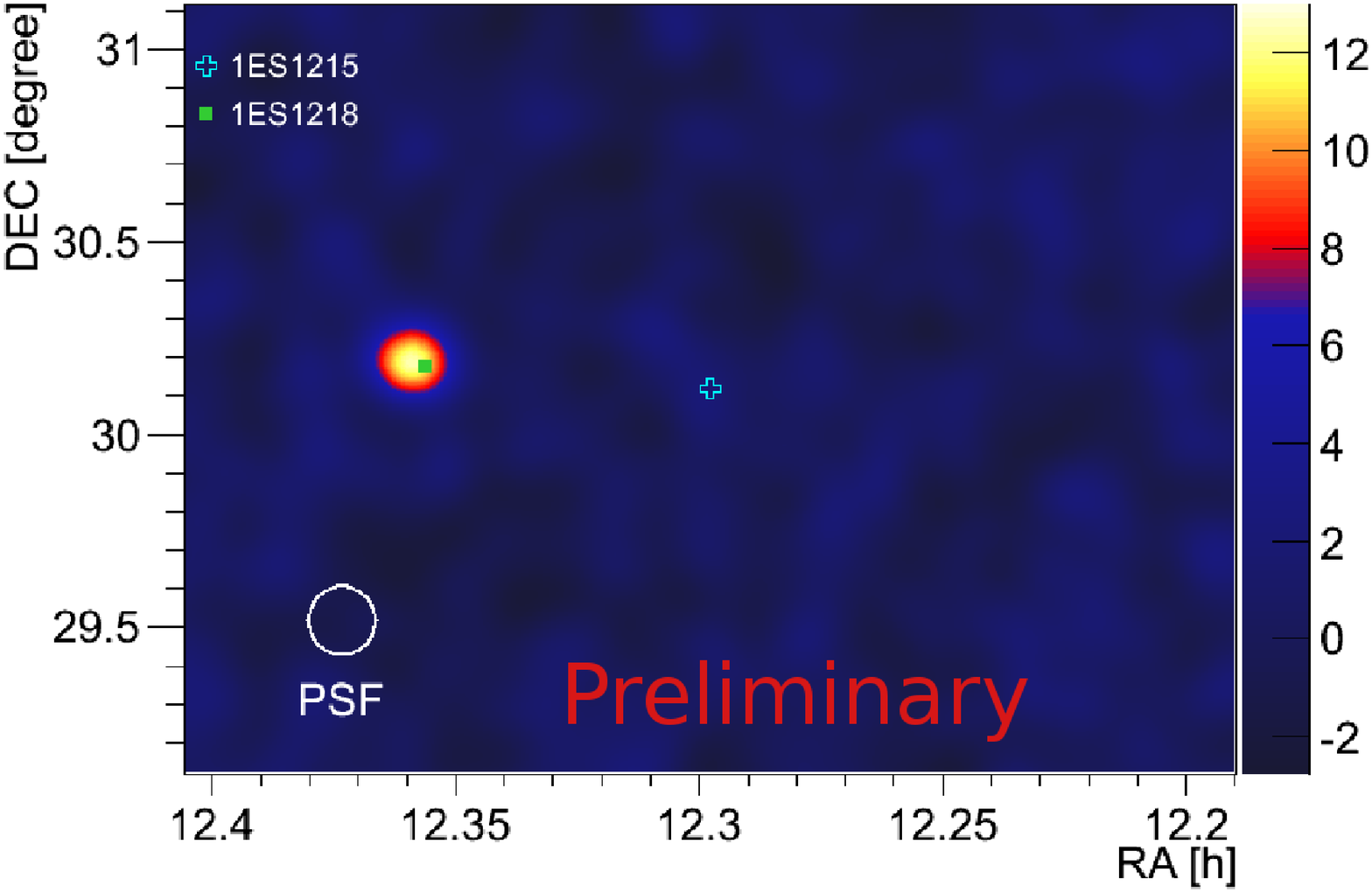}
\includegraphics[width=80mm]{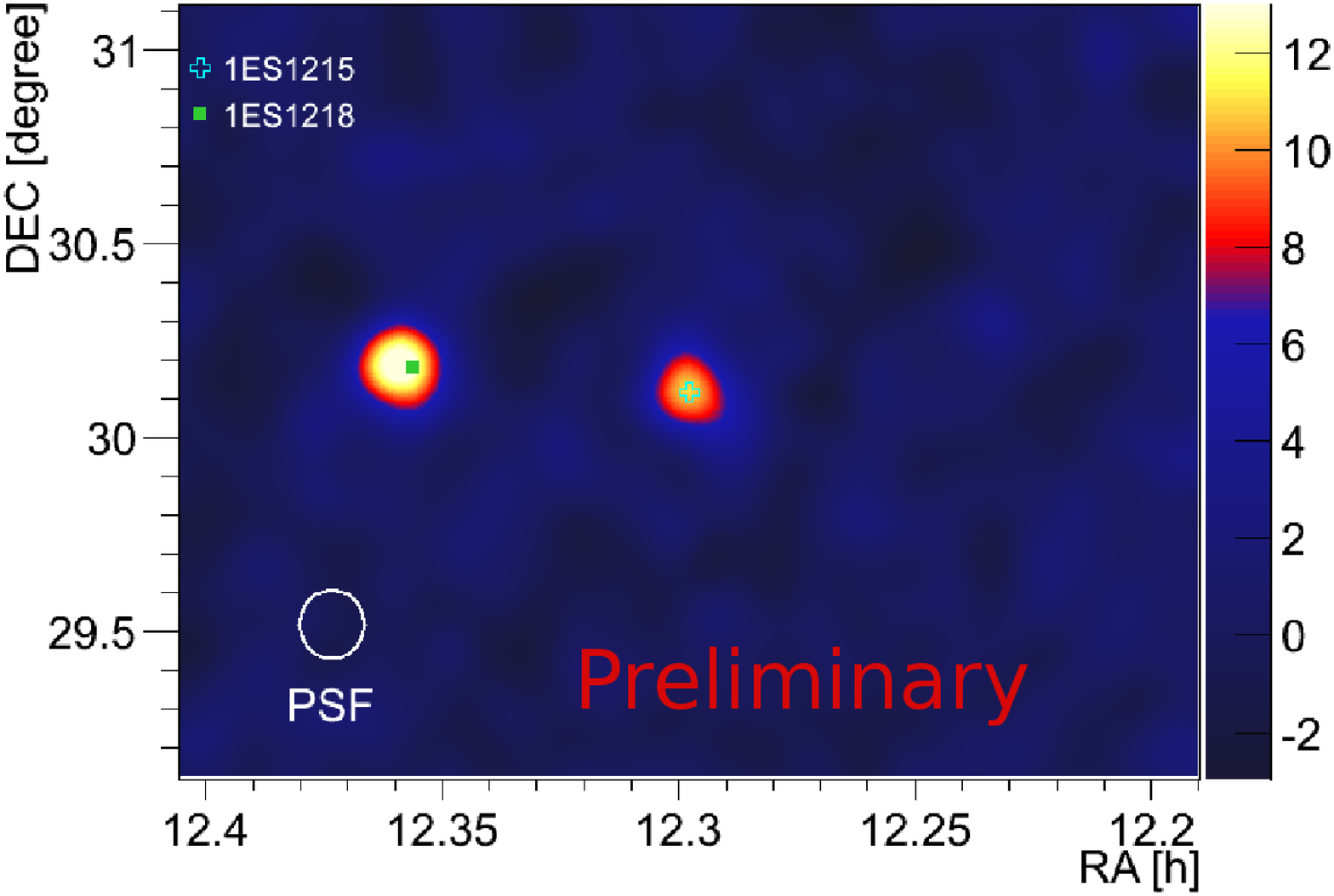}
\caption{Skymaps corresponding to 2010 observations (top) and 2011 observations (bottom). 
1ES1218+304 is clearly visible in both epochs while 1ES1215+303 is detected only in 2011.}
\label{skymaps}
\end{figure}
1ES~1215+303 and 1ES~1218+304 were observed by the MAGIC telescopes in 
January--February~2010, May--June~2010, and January--February~2011, for 
a total observation time of 48~hr. The observations 
were carried out in the so-called wobble mode~\cite{wobble} around 1ES~1215+303, 
with four pointing positions $0.4^{\circ}$ away from this source. 
1ES~1218+304 was not the primary target of these observations, but it was
always inside the MAGIC camera field of view, allowing its study at VHE as well. 
The data were taken in dark night and moderate moon conditions, and at zenith 
angles between 1 and 40 degrees.\\
For the analysis, the data were divided into two samples corresponding to 
two observing epochs, i.e. 2010 and 2011. The data were analyzed 
using the standard MAGIC software with additional adaptations incorporating the 
stereoscopic observations~\cite{lombardi}. The preliminary analysis of the 2010 data 
above an energy threshold of $\sim$300~GeV resulted in 3$\sigma$ significance level 
(computed using eq. (17) of Li$\&$Ma~\cite{LiMa}) for 1ES~1215+303, while 1ES~1218+304 
was detected at 12$\sigma$. In the 2011 data, the preliminary analysis detected a number 
of excess events of $N_{exc}(>300~\mbox{GeV})=110.4\pm13.6$, corresponding to a 
significance of $10.15\sigma$, which is the first significant detection of VHE 
$\gamma$-rays from 1ES~1215+303~\cite{Atel}. The $\theta^2$ distribution 
(the distribution of the squared angular distance between the arrival direction 
of the events and the nominal source position) is shown in Figure~\ref{theta2}.
The skymaps above $\sim$300~GeV corresponding to the 2010 and 2011 observations
are reported in Figure~\ref{skymaps}. As shown, 1ES~1218+304 is detected at a 12$\sigma$ 
significance level in both epochs, while 1ES~1215+303 is detected only in the 2011 data.
\section{Discussion and Outlook}
\begin{figure}
\includegraphics[width=50mm,angle=270]{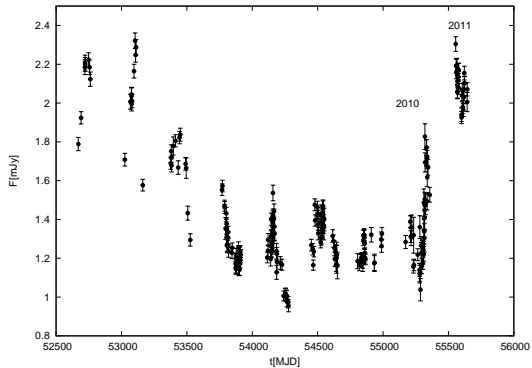}
\caption{Long-term optical light curve of 1ES~1218+304 from the Tuorla blazar monitoring program. 
Also 1ES~1218+304 was in a high optical state in January 2011.}
\label{opticalLC1218}
\end{figure}

The optical high state and the discovery of VHE $\gamma$-rays from 1ES~1215+303 
triggered multiwavelegth observations of the source. Multiwavelength data 
simultaneous and quasi-simultaneous to the MAGIC observations of 1ES~1215+303 were 
collected from radio to the $\gamma$-ray regime, including Mets\"ahovi 37~GHz data, 
optical R-band and polarization data from KVA, X-rays from Swift, and High Energy 
$\gamma$-rays from the \emph{Fermi}--LAT. The collected dataset allows us, for the first 
time, to construct a quasi-simultaneous SED of the source from radio to VHE $\gamma$-rays, 
and to study the connections between different wavebands. The analysis and interpretation 
of these data are still ongoing, but it is already evident that 1ES~1215+303 was in a high 
state from optical to the VHE $\gamma$-ray regime in January 2011 compared to previous 
observations from 2010. This demonstrates, once again, that optical monitoring data, 
which is easily obtainable, can be used for a successful triggering of VHE 
$\gamma$-ray observations.\\
For 1ES~1218+304 we will also study the long-term variability behavior in optical 
(see Figure~\ref{opticalLC1218}) and $\gamma$-ray bands. The simultaneous spectra
from \emph{Fermi}--LAT and MAGIC covers continuously more than 3 orders of
magnitude in energy. The study is ongoing, and the spectra can possibly bring 
new constraints on the EBL and on the intergalactic magnetic field.\\
The dedicated publications on these two sources are in preparation.

%




\begin{acknowledgments}
We would like to thank the Instituto de Astrof\'{\i}sica de
Canarias for the excellent working conditions at the
Observatorio del Roque de los Muchachos in La Palma.
The support of the German BMBF and MPG, the Italian INFN, 
the Swiss National Fund SNF, and the Spanish MICINN is 
gratefully acknowledged. This work was also supported by 
the Marie Curie program, by the CPAN CSD2007-00042 and MultiDark
CSD2009-00064 projects of the Spanish Consolider-Ingenio 2010
programme, by grant DO02-353 of the Bulgarian NSF, by grant 127740 of 
the Academy of Finland, by the YIP of the Helmholtz Gemeinschaft, 
by the DFG Cluster of Excellence ``Origin and Structure of the 
Universe'', by the DFG Collaborative Research Centers SFB823/C4 and SFB876/C3,
and by the Polish MNiSzW grant 745/N-HESS-MAGIC/2010/0.
\end{acknowledgments}

\bigskip 

\end{document}